\documentclass[aip,reprint]{revtex4-1}

\draft 
\usepackage{color}

\newcommand{\revise}[1]{{\color{black}{#1}}}
\usepackage{graphicx}
\newcommand{\ks}[1]{#1}

\begin{document}

\title{Study of magnetic reconnection at low-$\beta$ using laser-powered capacitor coils}

\author{H. Ji}
\email[]{Invited Speaker; email address: hji@princeton.edu}
\affiliation{Department of Astrophysical Sciences, Princeton University, Princeton, New Jersey 08544, USA}
\affiliation{Princeton Plasma Physics Laboratory, Princeton University, Princeton, New Jersey 08543, USA}
\author{L. Gao}
\affiliation{Princeton Plasma Physics Laboratory, Princeton University, Princeton, New Jersey 08543, USA}
\author{G. Pomraning}
\affiliation{Department of Astrophysical Sciences, Princeton University, Princeton, New Jersey 08544, USA}
\author{K. Sakai}
\affiliation{National Institute for Fusion Science, Toki-city, Gifu 509-5292, Japan}
\author{F. Guo}
\affiliation{Los Alamos National Laboratory, Los Alamos, New Mexico 87545, USA}
\author{X. Li}
\affiliation{Dartmouth College, Hanover, New Hampshire 03755, USA}
\author{A. Stanier}
\affiliation{Los Alamos National Laboratory, Los Alamos, New Mexico 87545, USA}
\author{A. Milder}
\affiliation{Laboratory for Laser Energetics, University of Rochester, Rochester, New York 14623, USA}
\author{R.F. Follett}
\affiliation{Laboratory for Laser Energetics, University of Rochester, Rochester, New York 14623, USA}
\author{G. Fiksel}
\affiliation{Center for Ultrafast Optical Science, University of Michigan, Ann Arbor, Michigan 48109, USA}
\author{E.G. Blackman}
\affiliation{Department of Physics and Astronomy, University of Rochester, Rochester, New York 14627, USA}
\affiliation{Laboratory for Laser Energetics, University of Rochester, Rochester, New York 14623, USA}
\author{A. Chien}
\affiliation{Department of Astrophysical Sciences, Princeton University, Princeton, New Jersey 08544, USA}
\author{S. Zhang}
\affiliation{Department of Astrophysical Sciences, Princeton University, Princeton, New Jersey 08544, USA}

\date{\today}

\begin{abstract}
Magnetic reconnection is a ubiquitous fundamental process in space and astrophysical plasmas that rapidly converts magnetic energy into some combination of flow energy, thermal energy, and non-thermal energetic particles. Over the past decade, a new experimental platform has been developed to study magnetic reconnection using strong coil currents powered by high power lasers at low plasma beta, typical conditions under which reconnection is energetically important in astrophysics. KJ-class lasers were used to drive parallel currents to reconnect MG-level magnetic fields in a quasi-axisymmetric geometry, similar to the Magnetic Reconnection Experiment or MRX, and thus this platform is named micro-MRX. This presentation summarizes two major findings from micro-MRX: direct measurement of accelerated electrons and observation of ion acoustic waves during anti-parallel reconnection. The angular dependence of the measured electron energy spectrum and the resulting accelerated energies, supported by particle-in-cell simulations, indicate that direct acceleration by the out-of-plane reconnection electric field is at work. Furthermore, a sudden onset of ion acoustic  bursts has been measured by collective Thomson scattering in the exhaust of magnetic reconnection, followed by electron acoustic bursts with electron heating and bulk acceleration. These results demonstrate that the micro-MRX platform offers a novel and unique approach to study magnetic reconnection in the laboratory in addition to the capabilities provided by traditional magnetized plasma experiments such as MRX and the upcoming FLARE (Facility for Laboratory Reconnection experiments). Future approaches to study other particle acceleration mechanisms and ion acoustic waves from magnetic reconnection are also discussed.
\end{abstract}

\maketitle

\section{Introduction}\label{intro}

Magnetic reconnection~\cite{yamada10,ji22} efficiently converts magnetic energy to plasma energy in the form of bulk flow, thermal \revise{particle}, and non-thermal particles through alteration of magnetic field topology. Magnetic reconnection occurs throughout the Universe~\cite{ji11} often as part of explosive phenomena such as solar flares and Earth's magnetospheric substorms. Magnetic reconnection has been confirmed to occur in the near-Earth space plasmas and in the laboratory plasmas via \textit{in-situ} measurements of local electromagnetic fields and plasma particles in and near the diffusion regions around the X-line \revise{where magnetic field lines change their connectivity}. Magnetic reconnection has been also long considered to occur in solar and more distant astrophysical plasmas via remote-sensing measurements of enhanced global morphological emission and \textit{ex-situ} measurements of accelerated particles. There has been substantial progress over the past 70 years in understanding the rate of magnetic reconnection in nearly collisionless plasmas. In  theoretical and numerical research~\cite{birn01} this includes identifying the electron dynamics responsible for breaking field lines, corroborated via space \textit{in-situ} measurements~\cite{burch16,torbert18}.  The associated kinetic structures and some energy conversion processes have also been identified via laboratory experiments~\cite{ji23b}. However, many open questions about the magnetic reconnection problem remain.

The major scientific challenges and research opportunities to meet them have been summarized in a number of recent community whitepapers~\cite{ji20,ji23a} submitted to several recent decadal surveys. Ten major problems are listed below:
\begin{enumerate}
\item \textit{Multiple scale problem}: How does reconnection couple global fluid (magnetohydrodynamic or MHD) scales to local dissipation (kinetic) scales? 
\item \textit{3D problem}: How does reconnection take place in 3D on global and local scales?
\item \textit{Energy problem}: How are particles heated and accelerated?
\item \textit{Boundary problem}: How do boundary conditions affect the reconnection process?
\item \textit{Onset problem}: How does reconnection start?
\item \textit{Partial ionization problem}: How does partial ionization affect reconnection?
\item \textit{Flow-driven problem}: What role does reconnection play in dynamos of flow-driven plasmas?
\item \textit{Turbulence and shock problem}: What role does reconnection play in  turbulence, collisionless shocks, and plasma transport?
\item \textit{Explosive phenomena problem}: How and and under what conditions does magnetic reconnection drive or 
follow explosive phenomena such as solar flares and coronal mass ejections? 
\item \textit{Extreme condition problem}: How does reconnection operate in extreme conditions such as intense radiation and relativity in astrophysics?
\end{enumerate}

Each of these interconnected problems can be tackled by a combination of theory, simulation, observation, and laboratory experiment. The latter has been important for testing theoretical and numerical predictions, to confirm observational evidence, and to discover new physics. Reference \citep{ji23b} concisely summarizes important results obtained over the past two decades in well-controlled and well-diagnosed laboratory experiments on collisionless reconnection. The experimental campaigns for these results, however, were performed on basic magnetized plasma facilities that  include devices specifically designated to study reconnection. In contrast, reconnection experiments performed in high-energy-density (HED) plasmas, powered either by lasers~\cite{nilson06,li07,willingale10,zhong10,fiksel14,zhong16,raymond18,fox20,fiksel21,fox21,ping23,pei16,yuan18,chien19,chien23,yuan23,zhang23} or by pulsed power~\cite{hare17,datta24b}, are more recent and less developed due to diagnostic and control difficulties, but have incurred rapid progress.

Many of these experiments in HED plasmas have used  platforms based on colliding plasma plumes and focus on the flow-driven regimes at high plasma $\beta_{up} \equiv (n_e T_e + n_i T_i)/(B^2_{up}/2\mu_0)\gg 1$ in the upstream \revise{region} of the magnetic reconnection site~\cite{nilson06,li07,willingale10,zhong10,fiksel14,zhong16,hare17,raymond18,fox20,fiksel21,fox21,ping23,datta24b}. Here $n_e$ and $n_i(=n_e/Z)$ are electron and ion number densities, respectively, \revise{and} $Z$ is ion charge. $T_e$ \revise{and} $T_i$ are the upstream electron and ion temperatures. Note that only the reconnecting component of the magnetic field within the reconnection plane, $B_{up}$, is used to define $\beta_{up}$; the uniform guide field component in the out-of-plane direction, while important for the plasma dynamics, is not included in the definition because its energy is not tapped  in the conversion of magnetic to kinetic energy during reconnection.

On the other hand, a new class of experiments~\cite{pei16,yuan18,chien19,chien23,yuan23,zhang23} uses the platforms based on laser-powered capacitor coils and focuses on magnetically-driven regimes at low upstream plasma $\beta_{up}$ $(\ll 1)$. Most of the basic magnetized plasma experiments are also in the same regime, but with $Z$ typically $\sim 1$. In contrast, the capacitor coil platform typically offers $\beta_{up} \ll 1 $ and $Z \gg 1$ simultaneously. These special characteristics, combined with unique diagnostic capabilities explained below, provide  much needed advantages for the capacitor coil platform to be used to study particle acceleration in addressing the problem \#3 (Energy problem) and problem \#2 (3D problem) listed above.

In assessing the implications of the experiments,  it is important to realize that essentially all impulsive energetic astrophysical phenomena attributed to magnetic reconnection are driven magnetically at low $\beta_{up}$. Solar flares are a prominent example where the magnetic field is the only abundantly available energy source. For such circumstances, \textit{the upper limit} for the averaged energy increase per plasma particle by reconnection, $\Delta E$, can well exceed their initial thermal energy, $T \equiv T_e=T_i$ for simplicity,
\begin{equation}
    \frac{\Delta E}{T}=\frac{B^2_{up}}{2\mu_0}\frac{1}{(n_e+n_i)T}=\frac{1}{\beta_{up}}\gg 1.
\end{equation}
Therefore, substantial increase in particle energy is expected for magnetic reconnection with a low $\beta_{up}$.  However, exactly how the converted magnetic energy is partitioned between particle heating versus non-thermal acceleration remains unresolved. Non-thermal particle acceleration during magnetic reconnection has been long observed in space plasmas~\citep{oieroset02}, solar flares~\citep{masuda94,krucker10,chen20}, and is plausibly responsible for the observed gamma-ray flares from the distant Crab Nebula~\citep{tavani11,abdo11,kroon16}. It has been also intensively studied theoretically and numerically~\citep{dahlin20,guo20,li21}, but only previously studied indirectly in the laboratory~\citep{savrukhin01,klimanov07,dubois17}. 

Although both the basic magnetized plasma experiments and laser-powered capacitor coil experiments have low $\beta_{up}$, there is a significant advantage of the latter  for diagnosing non-thermal particle acceleration. While \textit{in-situ} measurements of non-thermal particles are impossible for laser experiments and also generally difficult for basic magnetized experiments due to short Debye length for electrostatic energy analyzers~\citep{na23}, laser experiments are ideal for  \textit{ex-situ} measurements, analagous to many astronomical observations. The \textit{ex-situ} measurements of electrons are generally difficult  in basic magnetized plasma experiments due to the presence of cold plasma, dense neutral gas, and complex coils and vacuum structures in the plasma edge. The advantage of laser-powered capacitor coil platforms motivated our first set of micro-MRX experiments~\citep{chien23} to be discussed in this paper. For comparison, key parameters and characteristics of each research platform of magnetic reconnection are listed in Table~\ref{tab:table1}.

\begin{table*}[]
    \centering
    \begin{tabular}{c|c|c|c|c}
    \hline
    Platform & EOVSA/STIX & MMS & MRX/FLARE & Capacitor Coil=micro-MRX \\
    \hline
     Location & Solar corona & Earth's magnetotail & Lab & Lab \\
     \hline
    Regime & e/ion & e/ion & e/ion & electron only \\
     \hline
     Ion & H & H & H & Cu$^{+}$ ($Z=18$) \\
     \hline
     Length Scale $L_0$ & $10^7$ m & $6\times 10^8$ m & $0.8\to 1.6 $ m & $1$ mm \\
     \hline
     $e^-$ Density $n_{e0}$ & $10^{15}\; \text{m}^{-3}$ & $3 \times10^5\; \text{m}^{-3}$ & $10^{19}\; \text{m}^{-3}$ & $10^{24}$ m$^{-3}$ \\
     \hline
     System Size $\lambda = {L_0}/{d_i}$ or ${L_0}/{\rho_s}$& $4\times10^7$ & $1.3\times 10^3$ & $1.5\times 10^2 \to 10^3$ & $\sim 5$\\
     \hline
     $T_e$ & 200 eV & $600$ eV & $10\to30$ eV & $400$ eV\\
     \hline
     Reconnecting magnetic field $B$ & $.02$ T & $20$\; nT & $0.03\to 0.15$\; T & $100$\; T\\
     \hline
     Plasma Beta $\beta$ & $\sim 0.004$ & $\sim 4$ & $\sim 0.1$ & $\sim 0.06$\\
     \hline
     Lundquist Number $S$ & $10^{13}$ & $4\times10^{15}$ & $3\times 10^3\to 10^5$ & 200\\
     \hline
     $e^-$ Mean Free Path & $\sim 10^6$\; m & $\sim 10^{15}$\; m & $\sim 5$\; cm & $\sim 100$\; $\mu$m\\
    \hline
    Control & No & No & Yes & Yes \\
    \hline
    \textit{In-situ} measurement & No & Yes & difficult & No \\
    \hline
    \textit{Ex-situ} measurement & Yes & No & No & Yes\\
    \hline
    \end{tabular}
    \caption{Comparisons of research platforms for magnetic reconnection, including typical solar corona and near Earth space environments~\citep{ji11}. Here $\rho_s$ is ion sound radius. The micro-MRX platform is able to achieve low-$\beta$ conditions similar to these space and solar regimes. The mean-free-path of electrons listed in the table is for the thermal electrons. For  non-thermal accelerated electrons at $\sim$50 keV (see below), the mean free path would be $\approx 1.5$ m, much longer than the system size, $L_0$. In addition, solar (and astrophysical) observatories like the Expanded Owens Valley Solar Array (EOVSA)~\citep{gary18,chen20} and the Spectrometer/Telescope for Imaging X-rays (STIX) on board of Solar Orbiter~\citep{krucker20} rely on \textit{ex-situ} measurements, which are available on the micro-MRX platform, while space missions like MMS (Magnetospheric MultiScale)~\citep{burch16} rely on \textit{in-situ} measurements, which are being developed~\citep{na23} on MRX and FLARE. \revise{In addition, a general advantage of HED platforms is the feasibility of collective Thomson scattering~\citep{froula11} which can be powerful in directly detecting important plasma waves, as exemplified in this paper, as well as measuring local plasma parameters.}
    }
    \label{tab:table1}
\end{table*}

At low $\beta_{up}$, current-driven kinetic instabilities can be triggered in the reconnecting current sheet. Assuming an electron flow much faster than the ion flow, $V_e\gg V_i$,  current-driven instabilities are favored at large normalized drift velocities between electrons and ions
\begin{equation}
    \frac{V_d}{V_s}=\frac{j}{en_eV_s}=\frac{B_{up}}{\mu_0en_e V_s\delta},
    \label{eq:drift}
\end{equation}
where $V_d$, is the ion-electron drift velocity, $j$ is the peak current density, $\delta$ is the current sheet thickness, $V_s$ is ion acoustic speed,
\begin{equation}
    V_s \equiv \sqrt{\frac{ZT_e+T_i}{M}}=\sqrt{\frac{(Z+1)T}{M}},
    \label{eq:acoustic}
\end{equation}
and $M$ is ion mass. In the classical electron-ion reconnection regime, $\delta$ is of the order of the ion skin depth, $d_i \equiv c/\omega_{pi}=\sqrt{M/\mu_0e^2n_eZ}$, where $c$ is the speed of light and $\omega_{pi}$ is the ion plasma angular frequency. It follows that Eq.~(\ref{eq:drift}) becomes simply
\begin{equation}
    \frac{V_d}{V_s}=\sqrt{\frac{2}{\beta_{up}}},
    \label{eq:drift2}
\end{equation}
so that large values arise for low $\beta_{up}$.  A similar dependence on $\beta_{up}$ arises for electron-only reconnection where $\delta$ can be taken as the electron skin depth, $c/\omega_{pe}$, where $\omega_{pe}$ is the electron plasma angular frequency.

The laser-powered capacitor coil platform offers an important advantage over the basic magnetized experiments for studying current-driven kinetic instabilities of magnetic reconnection at high $Z$ values, and specifically $Z \simeq 18$ for the copper plasma made from our targets. In the cases described here, Ion Acoustic Waves (IAWs)~\citep{coppi71,smith72,coroniti77,sagdeev79} are not subject to ion Landau damping since the ion acoustic speed, Eq.~(\ref{eq:acoustic}), is much faster than the ion thermal speed, $V_{th,i}=\sqrt{T_i/M}$, since $Z\gg 1$. IAWs have been long conjectured to be important in providing a local, current-dependent anomalous resistivity required to sustain Petschek-type~\citep{petschek64} fast reconnection~\cite{ugai77,sato79,kulsrud01,uzdensky03}, but rarely studied~\cite{gekelman84} due to ion Landau damping, which is strong when $ZT_e\sim T_i$ as in  typical basic magnetized plasma experiments.

In Sec.~\ref{setup}, the experimental setup and diagnostics are described, followed by description and discussion of experimental results on electron acceleration~\citep{chien23}, and ion and electron acoustic waves~\citep{zhang23} in Sec.~\ref{results}. Discussion and future prospects are provided in Sec.~\ref{summary}.

\section{Experimental setup: Micro-MRX}\label{setup}

The laser-powered capacitor coil platform is facilitated by recent advances in strong external magnetic field generation using laser irradiation of a metallic coil target~\citep{daido86,fujioka13,santos15,gao16,fiksel16,law16,thikhonchuk17,goyon17,peebles20, morita21,morita23}. Such targets  usually consist of two parallel metallic foils connected by a thin wire that is bent into different coil shapes for generating various magnetic field configurations. High-energy lasers pass through the entrance hole in the front foil and irradiate the foil in the back. Hot electrons are generated during the intense laser-foil interaction and escape from the back foil, building up an electrical potential between the two foils. This produces a large current flowing through the connecting coil and therefore strong magnetic field generation. 

\begin{figure*}[t]
\includegraphics[width=.9\textwidth]{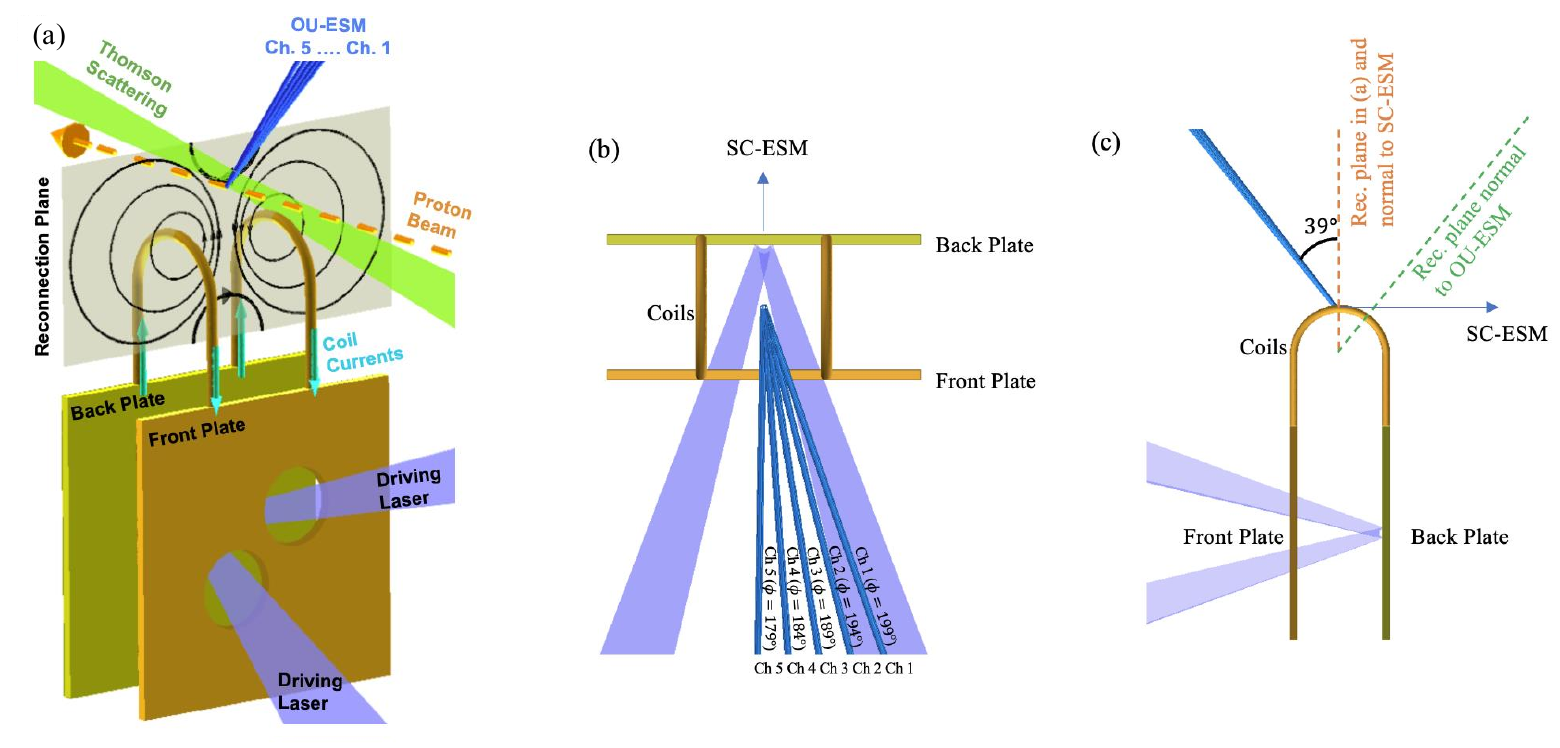}
\caption{(a) Experimental setup of magnetic reconnection experiments using a capacitor coil target with three major diagnostics. OMEGA-EP long-pulse beams pass through the front holes and irradiate the back plate delivering a combined $\sim 2.5$ kJ in $1$ ns. An electrostatic potential is induced between the (capacitive) plates and a large ($\sim 40-70$ kA) current is driven through the parallel U-shaped coils. The resulting magnetic fields undergo reconnection between the coils. One diagnostic is the OU-ESM, positioned $37.5$ cm away from the main interaction at an angle of $39^{\circ}$ away from the vertical. Five independent channels (directions indicated with the solid cyan lines above) are spaced $5^{\circ}$ apart, allowing  direct measurement of the angular spread of electrons in the azimuthal direction. The other diagnostics include a $2\omega$ ($527$ nm) Thomson scattering beam which probes the exhaust region $600$ $\mu$m above the center point at the top of the coils. The scattered light volume of $60\times 60\times 50$ $\mu m^3$ is collected by an $f/10$ reflective collection system. Ultrafast proton radiography probing through the reconnection plane is used to obtain magnetic field measurements. Due to the generation of the strong LPI signal by the short-pulse laser plasma interaction, ultrafast proton radiography is turned off to ensure a clean electron spectral measurement and its diagnostic port is fielded with the SC-ESM (aligned with the proton radiography line of sight and not shown in the figure). The SC-ESM is positioned at 9.5 cm away from the main interaction and perpendicular to the reconnection plane shown in (a). (b) Top-down view of the main target is shown, along with the OU-ESM channel orientation in the azimuthal direction and the SC-ESM orientation. (c) Side-on view of the main target shows the relative polar orientation of the OU-ESM channels and SC-ESM. The orange vertical dashed line represents the reconnection plane shown in (a) which is normal to the SC-ESM, and the green dashed line represents the reconnection plane that is normal to the OU-ESM line-of-sight.}
\label{fig:setup} 
\end{figure*}

The capacitor coil targets for the magnetically driven reconnection experiments discussed here consist of two connecting parallel coils~\citep{chien19} designed after successful measurement of the field generation by a single coil~\citep{gao16}. An example of the experimental schematic is shown in Fig.~\ref{fig:setup} using the OMEGA EP laser facility at the University of Rochester, NY. The targets were made from $50~\mu\textrm{m}$-thick Cu, and comprised of two square parallel plates with length $1.5~\textrm{mm}$ and two parallel U-shaped coils separated by $600~\mu\textrm{m}$. The U-shaped Cu coil, with a wire cross section of $50~\mu\textrm{m}\times100~\mu\textrm{m}$, had two 500-$\mu\textrm{m}$-long straight wires joined by a half-circular wire with a radius of curvature of 300 $\mu\textrm{m}$. The targets were fabricated by laser-cutting $50~\mu\textrm{m}$-thick sheet Cu, and then bending the coils into the desired shape. Target designs for other laser facilities in the US and abroad were  similar to those used on OMEGA EP, with slight modifications to accommodate laser configurations and optimize diagnostic output. Two EP long-pulse UV beams passed through the front holes and irradiated the back plate delivering a combined $\sim 2.5$ kJ in $1$ ns. \revise{The plasma in the coil region comes from the x-ray heated wires first as the preexisting plasma for magnetic reconnection, followed by the plasma from the ohmically heated wires. The diffusion of the plasma generated by the intense laser-foil interaction comes last, creating a situation of asymmetric reconnection in the downstream relevant to Earth's magnetotail and solar surface~\citep{ji22}. } 

In the double coil configuration, the voltage difference between the foils drives very similar currents in both coils, creating a quasi-axisymmetric magnetic reconnection geometry as a result of the anti-parallel magnetic fields in between the coils. The plasma between the coils is magnetized by the coil-driven anti-parallel magnetic field, forming a reconnection current sheet. This concept is very similar to that used for MRX, but at a much smaller scale, and the experiments are therefore named Micro-MRX. For MRX, a quadrupole magnetic configuration is formed by two flux cores, providing one public region where field lines wrap around both flux cores and two private regions where field lines wrap around only one flux core. Magnetic reconnection is induced thereby either increasing or reducing current in the flux cores by ``pushing" or ``pulling" flux between the public and  private regions, by charging capacitors to high voltages and then discharging~\citep{yamada97b}. For Micro-MRX, the current generated in the connecting coils increases while the laser pulse is on and decays after the laser pulse is turned off, providing  magnetically driven, quasi-axisymmetric ``push" and ``pull" reconnection.

Proton radiography was used to measure the current strength in the coils and any fine-scale structures in the reconnection region during reconnection~\citep{gao16}. This includes  ultrafast proton radiography using high energy protons generated by  target normal sheath acceleration (TNSA) and monoenergetic proton radiography using 14.7 MeV protons generated from D-$^{3}$He fusion inside an imploding capsule. The TNSA protons are broadband with energies up to 55-60 MeV~\citep{Gao2012,Gao2013} and provide a spatial resolution of 5 to 10 $\mu$m and a temporal resolution of a few picoseconds~\citep{Gao2015}. The mono-energetic protons provide $\sim$45-$\mu$m spatial resolution and $\sim$130-ps temporal resolution~\citep{Li2006}. As energetic protons probed the reconnection region, the proton beam spatial profile incurred variations from deflections by the Lorentz force, allowing inference of the field geometry.

Collective Thomson scattering was used to characterize  plasma parameters such as electron temperature and density and capture plasma waves in the reconnection region~\citep{Follett_RSI_2016}. In Micro-MRX experiments, a 527~nm probe laser was focused onto the plasma $600~\mu\textrm{m}$ above the X-line of the reconnection plane in Fig.~\ref{fig:setup}. The scattered light in a volume $60\times60\times50~\mu\textrm{m}^3$ was collected by an $f/10$ reflective collection system~\cite{katz12}, and the scattering angle was $63.4^{\circ}$. The collected scattered light was temporally and spectrally resolved by narrowband (7~nm window for ion-acoustic waves) and broadband (320~nm window for electron plasma waves) spectrometers, coupled with streaked cameras with a 5~ns streak window. 

Two time-integrated electron spectrometers -- the Osaka University electron spectrometer (OU-ESM) and the single channel electron spectrometer (SC-ESM) -- were used to measure the electron energy spectra. The OU-ESM is located $37.5~\textrm{cm}$ away from the coils at a polar angle of $39^{\circ}$, and scans an azimuthal range of $179^{\circ}-199^{\circ}$ with five equally-spaced detection channels [see Fig.~\ref{fig:setup}(b) and (c)]. The azimuthal angle spread was chosen to allow distinguishing between acceleration mechanisms in an axisymmetric setup, with symmetry in the polar direction. The SC-ESM is positioned at 9.5 cm away from the coil center and the measurement line-of-sight is perpendicular to the center reconnection plane as shown in Fig.~\ref{fig:setup}(a)). After reaching the spectrometer, an electron first passes through a pinhole $700~\mu\textrm{m}$ wide and $2~\textrm{cm}$ deep. Separation of electron energies is accomplished with permanent magnets placed along the detector line-of-sight, creating a magnetic field perpendicular to the line-of-sight. The Lorentz force deflects differently-energized electrons to different distances along the detector length onto an image plate. In general, impacts closer to the detector entrance represent lower-energy electrons. In the experiment, magnets were chosen corresponding to electron energies in the $20~\textrm{keV}-1~\textrm{MeV}$ range. During the electron spectral measurements, the proton radiography measurement was turned off to avoid contamination of the particle spectra due to laser-plasma instabilities (LPI).

\section{Experimental results}\label{results}

\subsection{Magnetic fields and reconnection signature}

\begin{figure}[h]
\includegraphics[width=.4\textwidth]{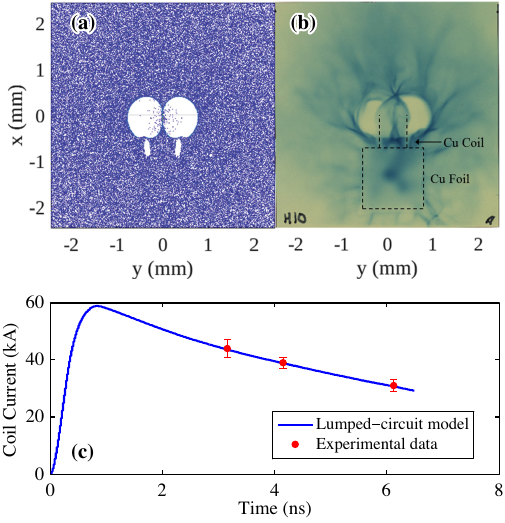}\\
\caption{Measurements of the coil currents using ultrafast proton radiography. (a) A synthetic radiograph of coil-generated magnetic fields using a coil current of $I=44$ kA and a proton energy of $E_{p}=24.7$ MeV, scaled to the target plane. (b) The experimentally measured proton radiograph at $t= t_{0} + 3.158$ ns for $E_{p}$ of 24.7 MeV. \revise{The color scale in (a) and (b) represents proton flux where darker regions in the images correspond to higher proton fluxes revealing proton accumulations due to deflections by the fields.} (c) Time evolution of the coil current as measured by the proton radiographs. Overlaid on top by the blue line is the coil current calculated by a lumped-circuit model with our experimental parameters. (a) and (b) are reproduced from Chien et al., Physics of Plasmas \textbf{26}, 062113 (2019), with the permission of AIP Publishing.}
\label{fig:current} 
\end{figure}

An example of the raw proton radiographs of the coil target after laser irradiation in the face-on radiography geometry is shown in Fig.~\ref{fig:current}(b). The data were obtained at $t= t_{0} + 3.158$ ns for 24.7 MeV protons generated by TNSA, where $t_{0}$ is the arrival time of the drive beams at the surface of the back Cu foil. The data show the rectangular Cu foil, the fiber stalk that holds the entire target, and the straight-part of the two U-shaped coils as viewed from the  face-on direction to the target. The dashed line overlaid on top is the contour of the original target in a face-on view, in good agreement with the experimental measurement. The primary features are the formation of two prolate voids generated by the magnetic fields of the driven coil currents, and a center flask-shaped structure between the voids that indicates the occurrence of magnetic reconnection. 

The amplitude of the coil current and the current-generated magnetic fields are estimated by matching the theoretically
calculated and measured proton voids in the proton radiographs. Synthetic radiographs are generated via a particle ray-tracing code using the same radiography geometry as in the experiments. The proton beam generated by TNSA propagates through the coil target. The amplitude and distribution of the three-dimensional magnetic fields are calculated using the Biot-Savart law. As the energetic proton beam traverses  the field region,  the beam's spatial profile varies from  Lorentz force deflections. Synthetic proton images are constructed by tracing each proton trajectory and counting the accumulated protons at the detector plane. Figure~\ref{fig:current}(a) shows the synthetic proton image with matching voids as measured in the experiment, indicating a coil current of $I=44$ kA at $t= t_{0} + 3.158$ ns. Electrical fields were found to be negligible in contributing to the proton features \citep{gao16}. 

Figure~\ref{fig:current}(c) shows the time evolution of the coil current measured by the proton radiographs at different probe times. The blue line overlaid on top is the coil current calculated by a lumped-circuit model using our experimental parameters~\citep{fiksel2016,chien21}. The coil current profile is approximated by a linear rise during the laser pulse, followed by exponential decay after laser turn-off at 1 ns. Extrapolating the experimentally measured current magnitude to $t=1$ ns, the maximum coil current is inferred to be $\sim57\pm4$ kA, which corresponds to a coil center magnetic field strength of $\sim110$ T and an upstream reconnection magnetic field strength of $50.7~\textrm{T}$. 

\begin{figure}[t]
\includegraphics[width=.45\textwidth]{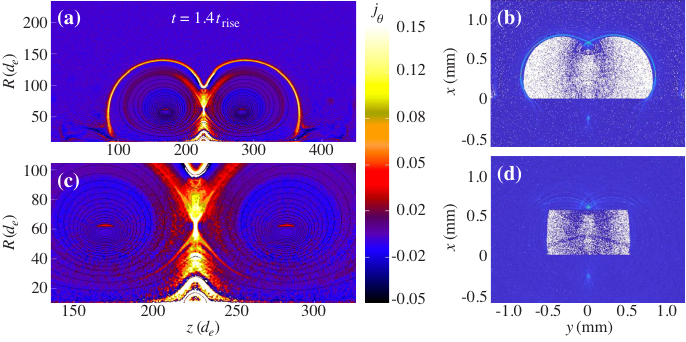}\\
\caption{The ``center feature'' in proton radiographs reproduced synthetically. (a) Full box and (c) zoomed-in 2D profiles in the $(R,z)$ plane of out-of-plane current $j_{\theta}$ in the normalized unit of $j_{0}=en_{e0}c$ with respective synthetic proton radiographs (b) and (d) for $t=1.4~t_{\textrm{rise}}$ where $t_{\textrm{rise}}$ is the current rising time of 1 ns. The synthetic proton radiographs are generated by ``sweeping'' the electromagnetic field structure in a semicircle, advancing a cone of protons by raytracing through the fields in the $z$ direction, and recording the projected proton positions in the $y$-$x$ plane. \revise{The color scale in (b) and (d) represents the proton flux where darker regions in the images correspond to higher proton fluxes revealing proton accumulations due to deflections by the fields.} The diamagnetic return current can be seen in the full $j_{\theta}$ profile, but not in the zoomed-in profile. Reproduced with permission from Chien et al., Nature Physics \textbf{19}, 254 (2023). Copyright 2023 Springer Nature.}
\label{fig:flask}
\end{figure}

A striking ``flask-like'' feature with high proton fluence is observed in between the two prolate voids. This feature is not present in synthetic radiographs simulated with only the coil magnetic fields [Fig.~\ref{fig:current}(a)]. Analytical calculations~\citep{chien19}, MHD simulations using the code FLASH~\citep{zhang23}, and PIC simulations using the code VPIC~\citep{chien23} were carried out.  All point to the conclusion that this center feature is formed by the out-of-plane current from magnetic reconnection. 

Figure~\ref{fig:flask} \revise{shows} results of the VPIC simulations. Details of the simulation setup can be found in Ref.~\citep{chien23}. Electromagnetic fields from the VPIC simulations of the experiments were prescribed in the particle ray tracing calculations to  generate  synthetic proton radiographs. Figure~\ref{fig:flask}(a) shows the out-of-plane current $j_{\theta}$ profiles at $t=1.4~t_{\textrm{rise}}$, where $t_{\textrm{rise}}$ is the current rise time of 1 ns. Strong push reconnection is seen at this time. The corresponding synthetic proton image is shown in Figure~\ref{fig:flask}(b) where the center feature is reproduced. Two primary features in the out-of-plane current $j_{\theta}$ profiles are potentially responsible for creating this center feature: the push reconnection current sheet and diamagnetic return current. To de-convolve the effects of each on the synthetic proton radiographs, raytracing is performed on a ``zoomed'' field, where the diamagnetic return current is largely shielded out [Fig.~\ref{fig:flask}(c)]. The center feature is maintained in this radiograph [Fig.~\ref{fig:flask}(c)], indicating the source of the center feature as the push reconnection current sheet. Thus, the presence of a similar center feature in experimental radiographs is indicative of push reconnection and the corresponding electromagnetic fields.

\subsection{Plasma parameters and reconnection regime}

\begin{figure}[t]
\includegraphics[clip,width=.45\textwidth]{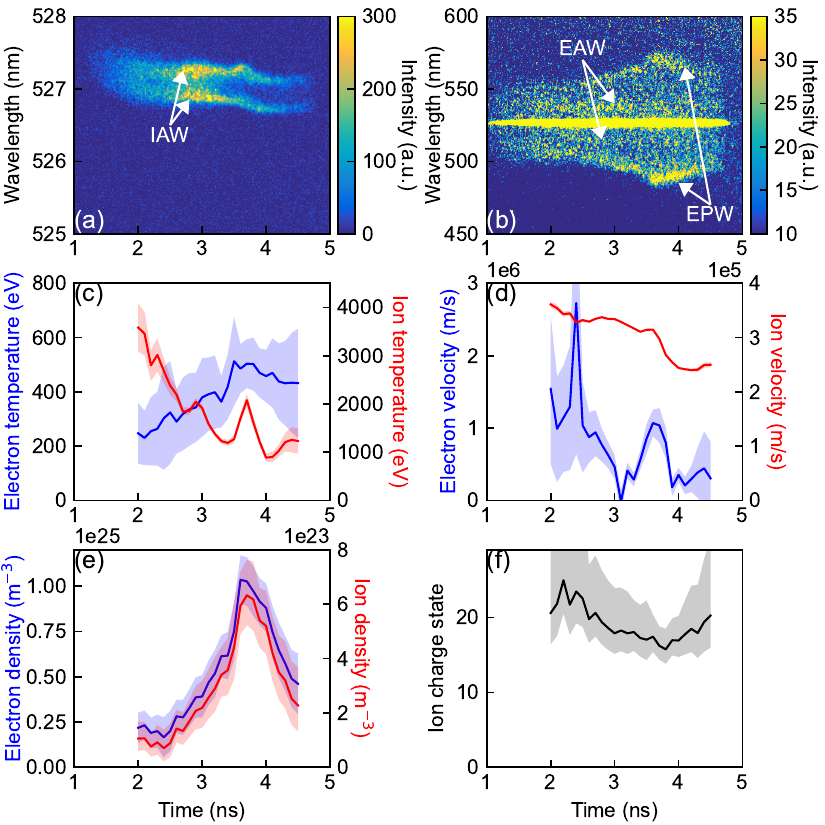}
\caption{Experimentally measured (a) IAW and (b) EPW spectra and inferred plasma parameters (c) temperature, (d) velocity, (e) density, and (f) \revise{ion charge state}. \label{fig:TStime}}%
\end{figure}

Collective Thomson scattering was used to measure plasma conditions in the reconnection region. Assuming Maxwellian distributions for ions and electrons, the plasma temperature ($T_i$ and $T_e$), plasma density ($n_i$ and $n_e$) and \revise{ion charge state} $Z$ can be determined from the measured spectra of ion acoustic waves (IAWs) and electron plasma waves (EPWs). Figures~\ref{fig:TStime}(a) and \ref{fig:TStime}(b) show temporally and spectrally resolved IAW and EPW spectra from collective Thomson scattering measured at 600 $\mu$m downstream of the central point between the top of the coils (see Fig.~\ref{fig:setup} for Thomson scattering setup), with a propagation direction within the reconnection plane and pointing $\sim$17$^\circ$ away from downstream. The measurement time was at 1 -- 5 ns with respect to the onset of the laser irradiation to the capacitor coil target. (See Sec. IIID for the analysis of Thomson spectra at a later time.) Figures~\ref{fig:TStime}(c)--\ref{fig:TStime}(f) show the time evolution of plasma temperature, velocity directed to the downstream, plasma density, and \revise{ion charge state} by forward fitting the synthetic IAW and EPW spectra respectively to the experimentally measured broadband spectra~\citep{froula11}. A single Maxwellian for electrons and ions was assumed. \revise{In the IAW analysis, the relative drift velocity between electrons and ions is set as a parameter to explain the asymmetry of two IAW peaks.} The analyses were performed every 0.1 ns, including the spatial gradients of density and velocity. \revise{The error bars correspond to the standard deviation of each parameter estimated by the least squares fitting.}

\begin{figure}
    \centering
    \includegraphics[width = .45\textwidth]{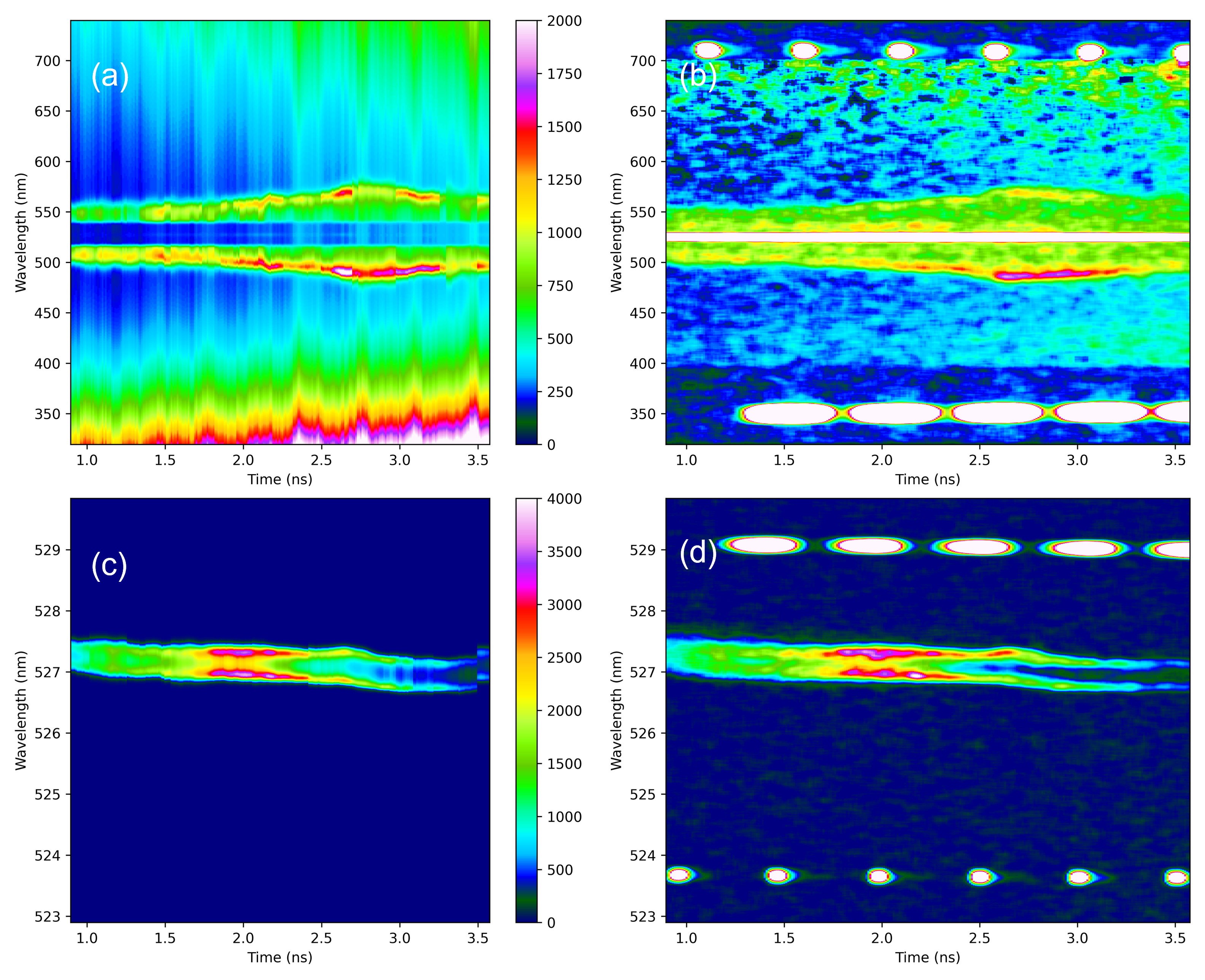}
    \caption{Raw EPW (b) and IAW (d) spectral data are shown compared with the matching calculated EPW (a) and IAW (c) from a simultaneous fit to the spectra. \revise{The calculated spectra an assemblage of individual fits at different times combined to form spectrum as a function of wavelength and time.}}
    \label{fig:TSADAR}
\end{figure}

To cross-benchmark the above analyses, Figure \ref{fig:TSADAR} shows the results of using the TSADAR code\citep{Milder_2024} to analyze the IAW and EPW spectra simultaneously from the same Thomson scattering data as Figure~\ref{fig:TStime}. \revise{The raw data is divided into individual lineouts representing the spectrum at a single temporal slice, the measured conditions are found by matching the Thomson scattering spectral model to the data and the best matching spectra are stacked temporally to produce the images Fig.~\ref{fig:TSADAR}(a) and (c). The data is only analyzed in the neighborhood of the data, i.e. 450-600 nm for the EPW and 526-528 nm for the IAW, ignoring the temporal calibration fiducials recorded at the top and bottom of the data images.} This analysis corroborated that of Figure \ref{fig:TStime} and checked for the influence of temporal gradients and super-Gaussian electron velocity distribution functions, as well as accounting for all instrumental effects such as finite aperture corrections. These considerations were found to have negligible effect on the inferred parameters in this regime. 

At 600 $\mu m$ downstream from the X-line, the nominal plasma parameters are: $Z\sim 18$, $n_e \sim 3\times 10^{24}$ m$^{-3}$, $n_i \sim 1.7\times 10^{23}$ m$^{-3}$, $T_e \sim $400 eV. This results in $\beta \approx \beta_e \sim 0.05$ for $B=100$ T, since $n_i \ll n_e$. This low-$\beta$ condition is favorable for studies of particle heating and acceleration, and also for IAW excitation since $Z\gg 1$ as discussed in Sec.~\ref{intro}. Note that the EPW spectra show EAW peaks that are explained by non-Maxwellian electron distribution functions \citep{zhang23}. As a result, the Lundquist number is large at $S=10^3-10^4$ but the system size $L=600$ $\mu$m is not too large compared to the ion skin depth due to heavy copper ions so the normalized size $\lambda=L/d_i \sim 1.4$. This places our current experimental setup in the electron-only regime~\citep{phan18} of the magnetic reconnection phase diagram~\citep{ji11,ji22}. In this regime, ions are unmagnetized by the reconnecting magnetic field but they can still respond to the resulting electric field if a sufficient time is given. It is this case in our setup where IAWs have been observed, see Sec.~IIID below.

\subsection{Electron acceleration}

\begin{figure}[h]
\includegraphics[width=.5\textwidth]{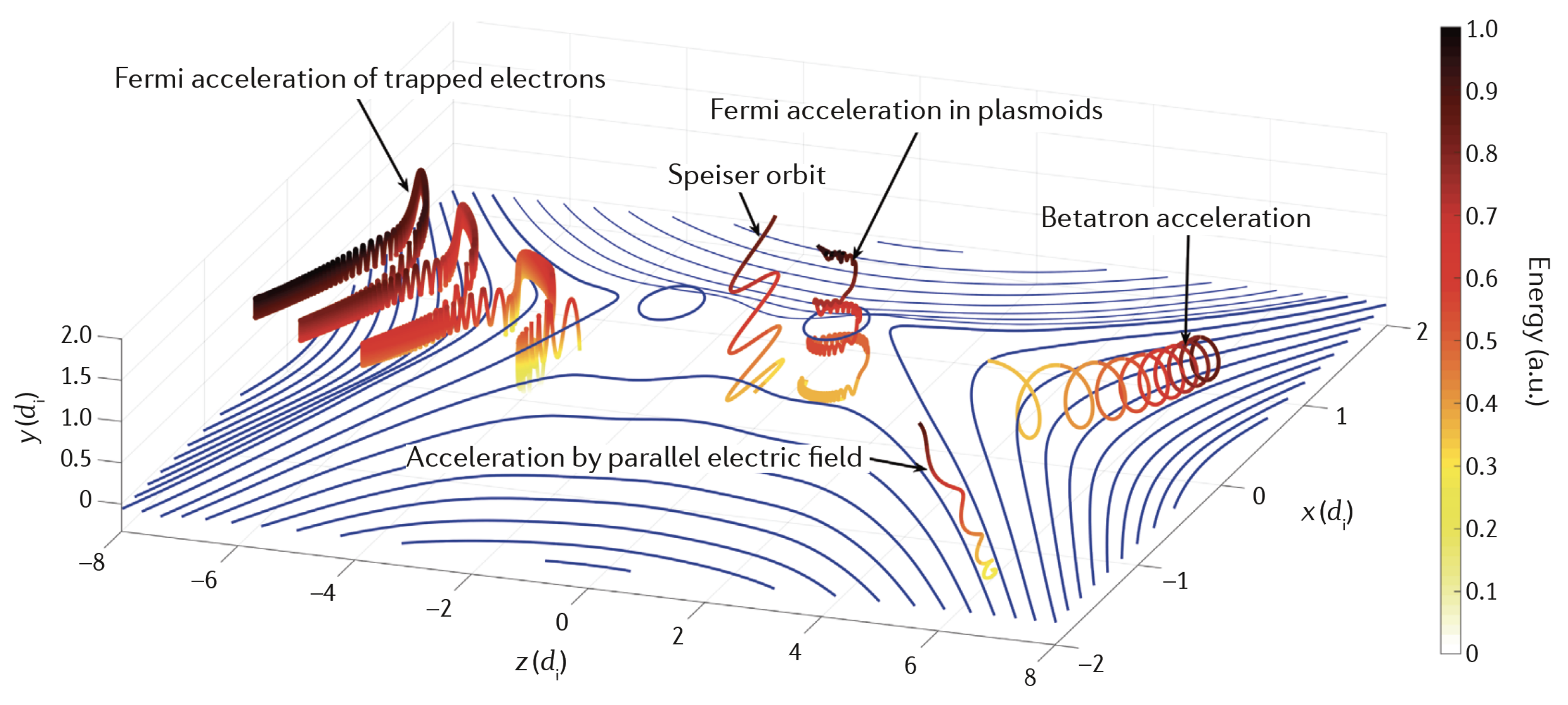}\\
\caption{Various proposed particle acceleration mechanisms during magnetic reconnection: direct acceleration by reconnection electric field while exhibiting Speiser orbit, parallel electric field acceleration, Fermi acceleration due to field line curvature in downstream or in plasmoids, and Betatron acceleration due to increasing magnetic field strength. Reproduced with permission from Ji et al., Nature Reviews Physics \textbf{4}, 263 (2022). Copyright 2022 Springer Nature.}
\label{fig:acceleration} 
\end{figure}

Theoretically and numerically, non-thermal particle acceleration has recently been studied extensively ~\citep{dahlin20,guo20}, and various acceleration mechanisms have  long been proposed~\citep{ji22}. These  include parallel electric field acceleration~\citep{egedal13}, Fermi acceleration~\citep{drake06}, betatron acceleration~\citep{hoshino01}, and direct acceleration by reconnection electric field~\citep{zenitani01,uzdensky11d}. Figure~\ref{fig:acceleration} illustrates these proposed particle acceleration mechanisms in and near the reconnection site. In the electron-only reconnection, electron heating and acceleration can be studied if electrons have sufficient
space and time to receive energy from the magnetic field,  as in our case.

Our primary diagnostic is the time-integrated measurement of electron energy spectra by the Osaka University Electron Spectrometer (OU-ESM). This is aimed nearly tangentially along the X-line with 5 channels at different azimuthal detection angles that range from 179$^\circ$ to 199$^\circ$ towards one of coils.  Channel \#5 has an angle of 179$^\circ$ and is most tangential to the X-line direction.

\begin{figure}[t]
\includegraphics[width=.4\textwidth]{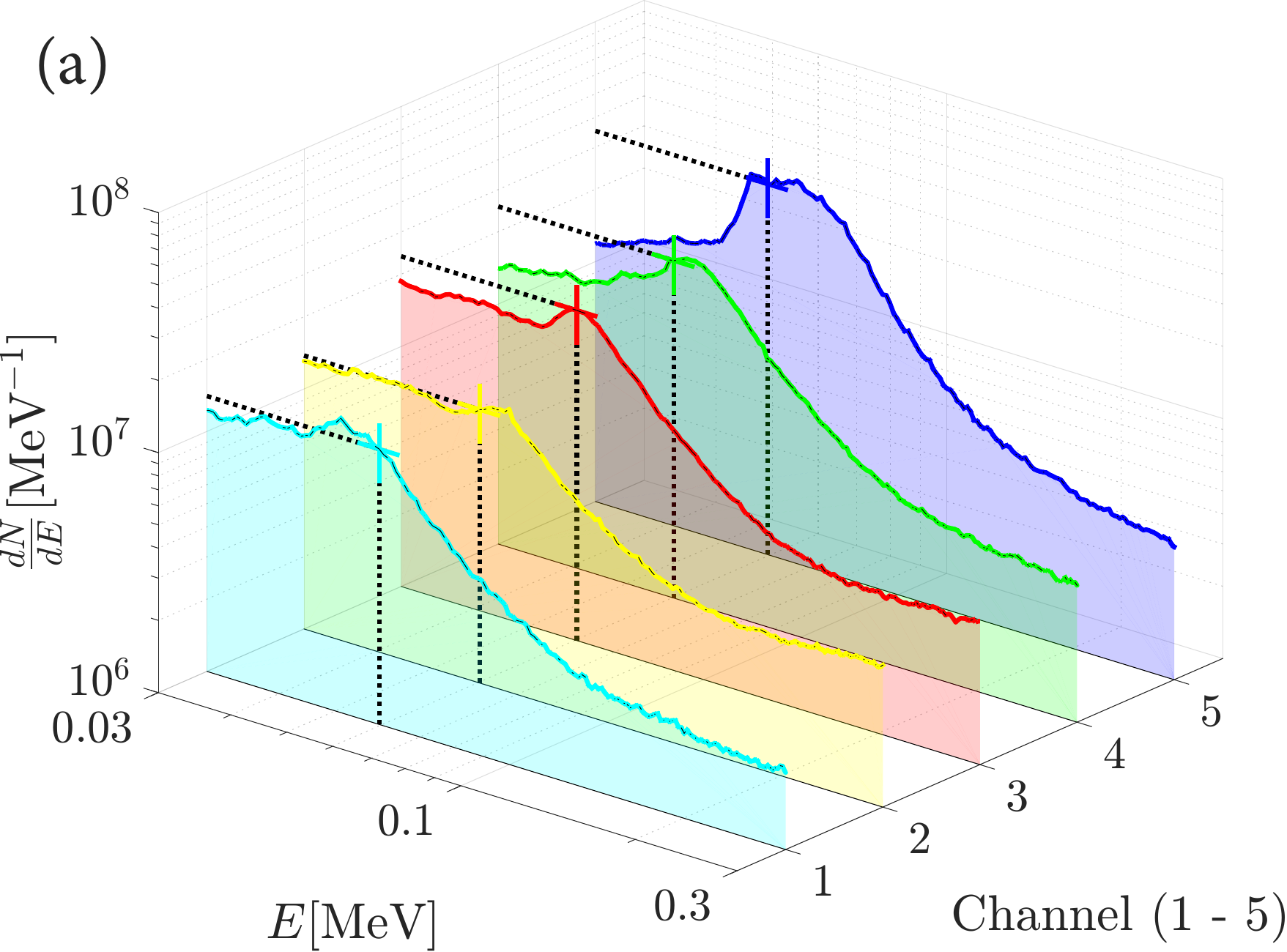}\\
\vspace{5mm}%
\includegraphics[width=.4\textwidth]{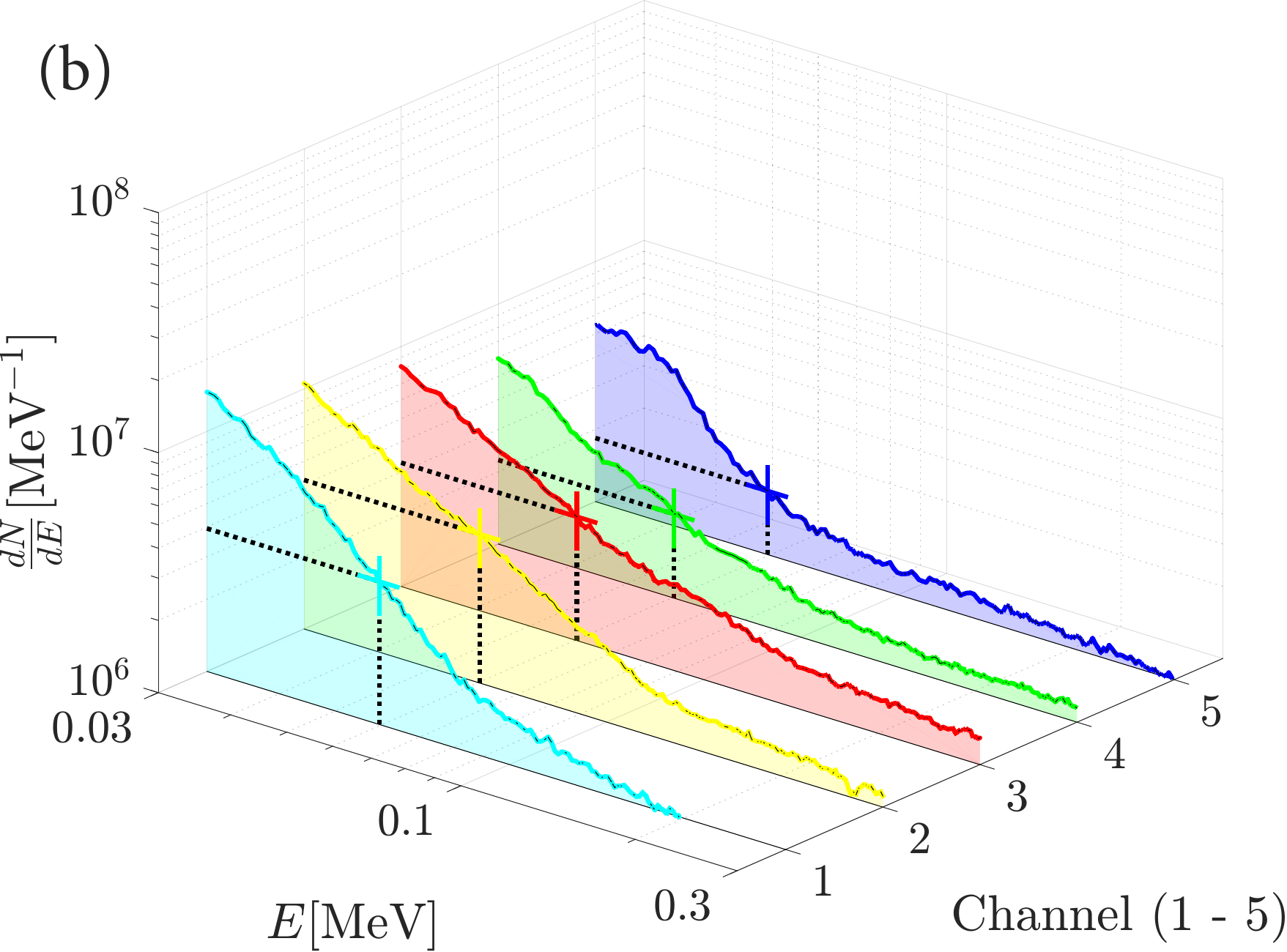}\\
\vspace{5mm}%
\includegraphics[width=.4\textwidth]{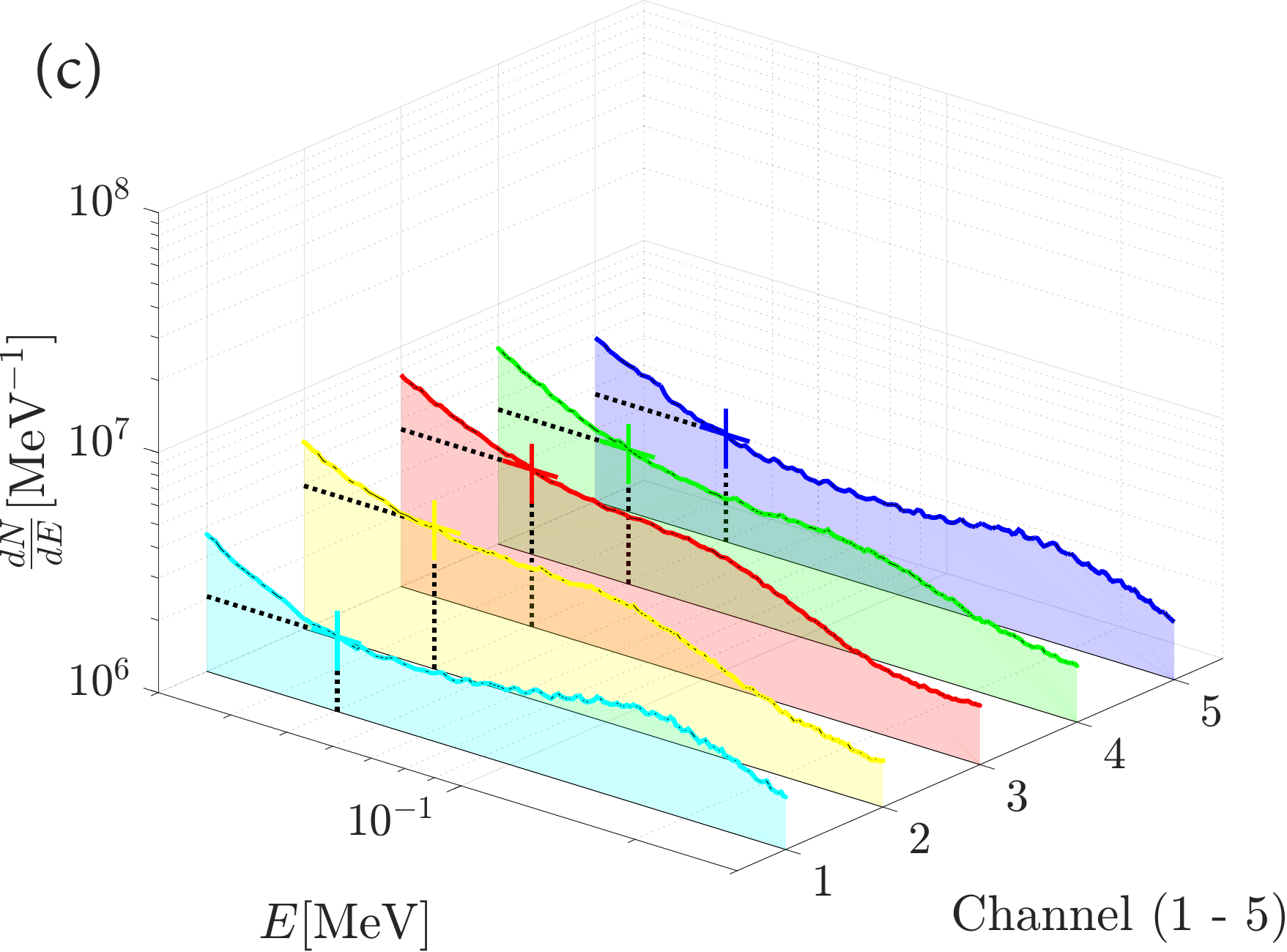}%
\caption{Electron energy spectra measured by 5-channel OU-ESM (a) by magnetic reconnection driven by two capacitor coils, (b) with only one capacitor coil, and (c) with no coils. Crosses at 60 keV represent the characteristic horizontal and vertical error bars. Reproduced with permission from Chien et al., Nature Physics \textbf{19}, 254 (2023). Copyright 2023 Springer Nature.
\label{fig:OU_ESM}}
\end{figure}

Figure~\ref{fig:OU_ESM} (a) exemplifies the measured electron energy spectra in the magnetic reconnection case with two capacitor coils. A peak in the energy range of 40-70 keV is identified, and it becomes clearest in the channel \#5 which measures along the most tangential direction to the reconnection X-line. That no such peaks exist in the control cases shown in Fig.~\ref{fig:OU_ESM} (b) with only one coil and Fig.~\ref{fig:OU_ESM} (c) with no coil establishes that the peak is due to reconnection.

The measured angular dependence suggests that the acceleration of non-thermal electrons is due to the reconnection electric field, $E_{rec}$, in the out-of-plane direction of the anti-parallel reconnection~\citep{zenitani01}. The required sign of $E_{rec}$  to be directly responsible for the observed acceleration points to the ``push'' reconnection driven by the increasing current in both coils, which is expected after lasers irradiate the capacitor coil target but possibly with some delays suggested by matching simulation (see below).

\begin{figure}[t]
\includegraphics[width=.35\textwidth]{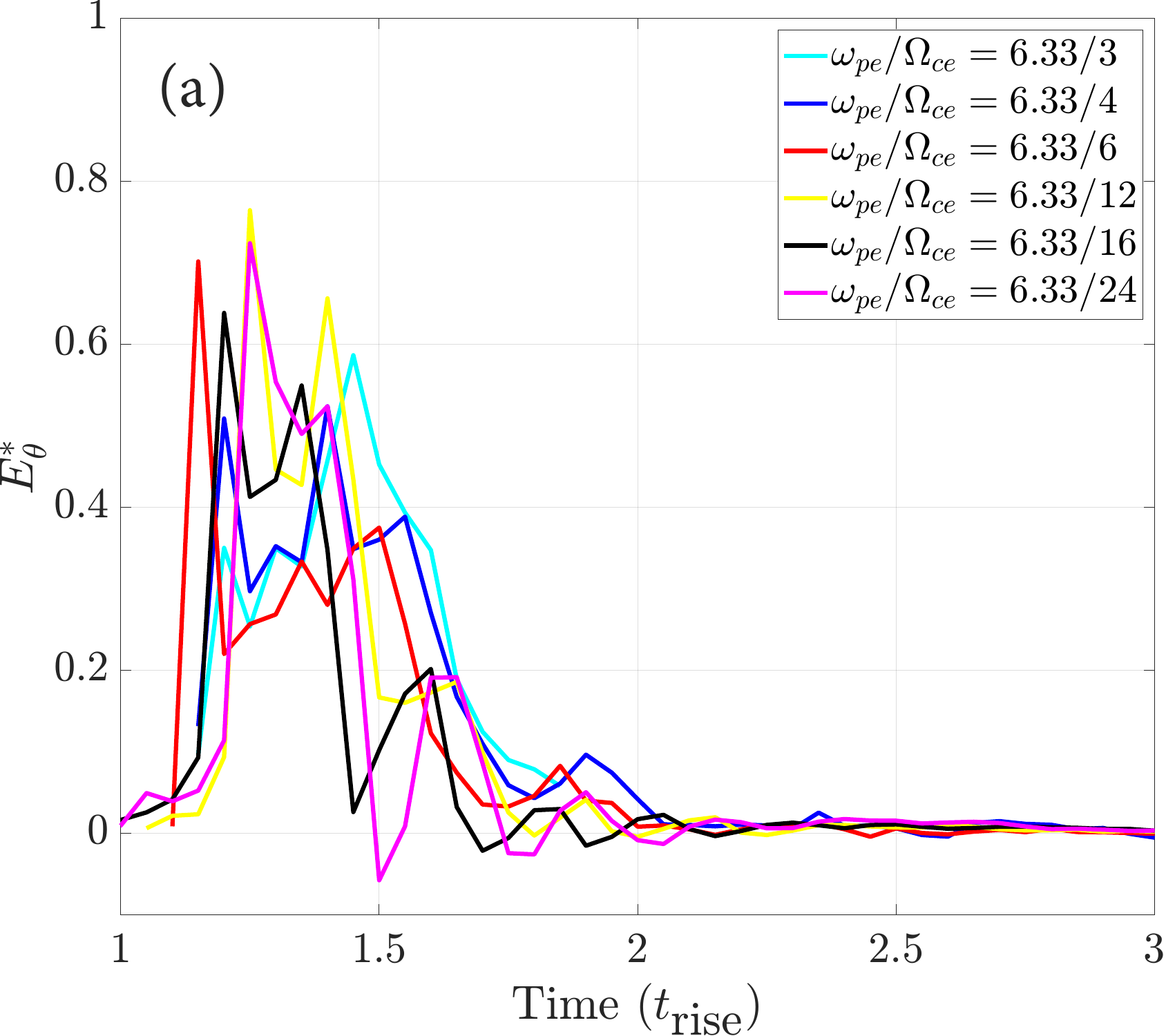}\\
\vspace{5mm}%
\includegraphics[width=.35\textwidth]{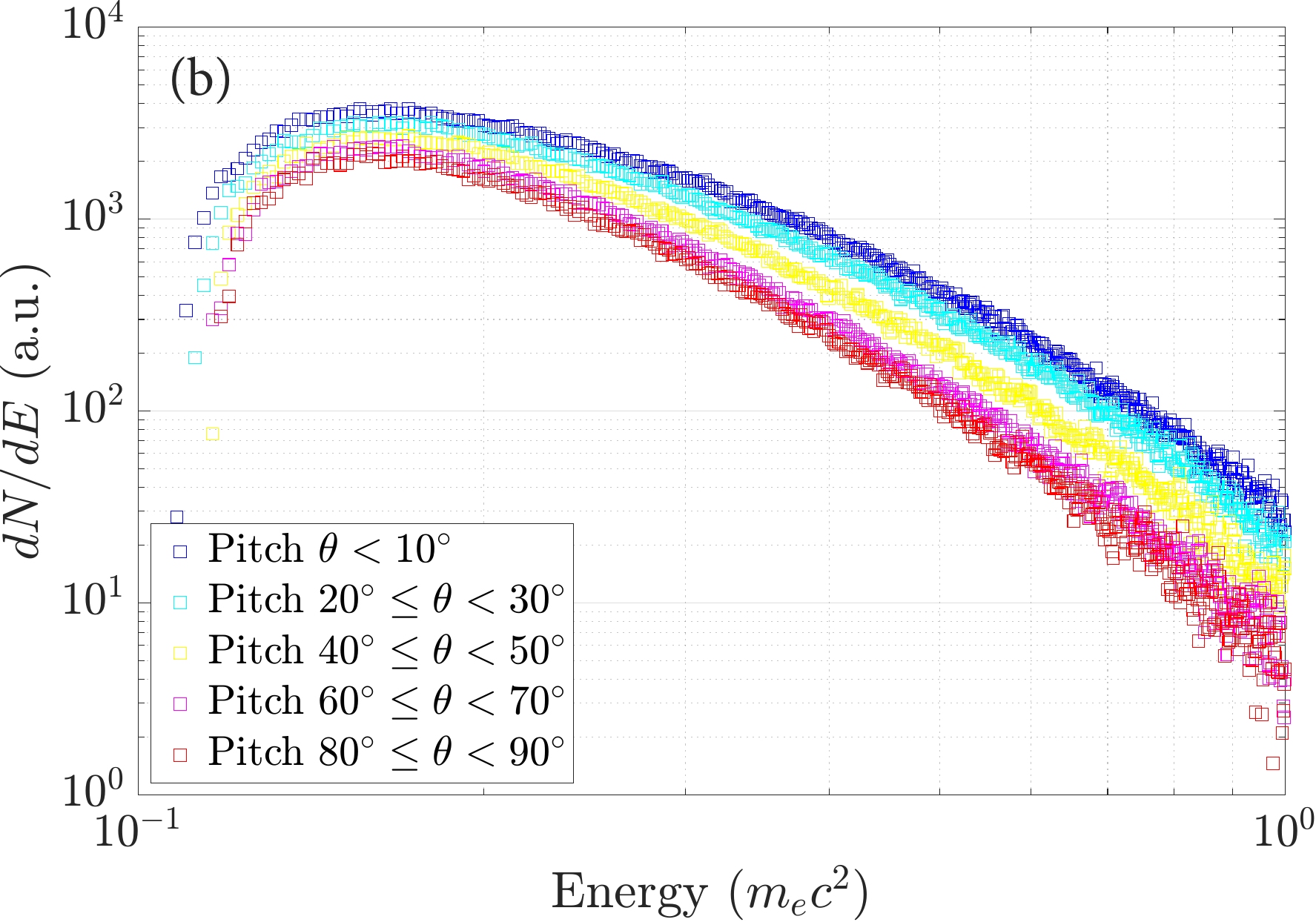}
\caption{Normalized reconnection rate, $E_\theta^*\equiv\alpha$, as a function of time (a) and non-thermal difference energy spectra of electrons in the central area between coils after reconnection compared with the baseline before reconnection (b). Reproduced with permission from Chien et al., Nature Physics \textbf{19}, 254 (2023). Copyright 2023 Springer Nature.
\label{fig:VPIC}}
\end{figure}

The magnitude of $E_{rec}$ can be estimated as $E_{rec}=\alpha V_A B_0$ where $\alpha$ is the reconnection rate,  $B_0$ is the upstream reconnecting magnetic field of $50.7$ T (see Sec.~\ref{setup}), and the Alfv\'en speed $V_A=B_0/\sqrt{\mu_0n_i M}$ where here $n_i$ is the ion density at the X-point. The reconnection rate $\alpha$ is typically $\sim 0.1$ for electron-ion reconnection but
for electron-only reconnection it is a function of the normalized size~\citep{pyakurel19}.  For the small size limit that applies to  our case, $\alpha \simeq 0.6$. Therefore, for a range of $n_e=(1-5) \times 10^{24}$ m$^{-1}$, we estimate $E_{rec}=(1.3-3.0)\times 10^7 $ V/m. If we take the characteristic acceleration distance $d=1000\mu$m in the out-of-plane direction, the accelerated electrons by the reconnection electric field should have an energy of $E_{rec}d = 13-30$ keV, which is  within a factor of 2 of the measured energy of $30-70$ keV for the spectral peak. We consider this consistent with our interpretation given the uncertainties  in the relevant parameters.

To further test the above interpretation, a series of axisymmetric Particle-In-Cell simulations~\citep{chien23} were performed using \revise{the} VPIC (Vector PIC) code~\citep{bowers08}. Due to compuational resource limitations, choices were made to achieve a real mass ratio of copper ions to electrons as well as a realistic plasma $\beta$. This meant using an artificially stronger magnetic field and  hotter plasma,  reducing the ratios of electron plasma frequency to gyrofrequncy, $\omega_{pe}/\omega_{ce}$.

Figure~\ref{fig:VPIC} (a) shows the reconnection rate $\alpha$ as a function of time measured in the reconnection region, indicating that the peak $\alpha \sim 0.6$ is approximately confirmed when $\omega_{pe}/\omega_{ce}$ approaches  the realistic value of 6.33. As a consequence of reconnection, the accelerated electron energy spectra shown in Fig.~\ref{fig:VPIC} (b) exhibits an angular dependence qualitatively consistent with the experimentally measured spectra shown in Fig.~\ref{fig:OU_ESM}(a). The energy of the spectral peak is also consistent with the experimentally measured range of 40-70 keV.

\begin{figure}[t]
\includegraphics[width=0.35\textwidth]{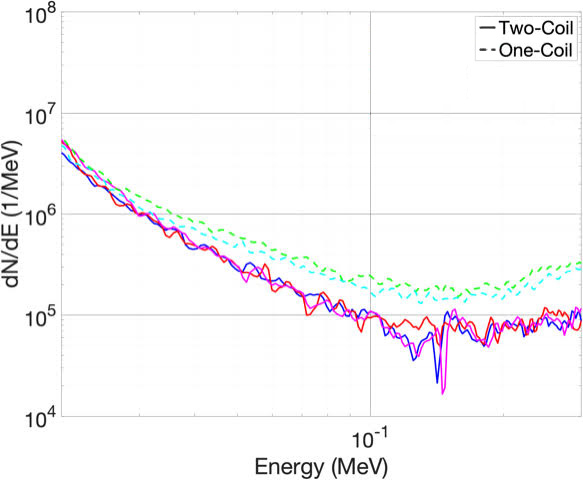}
\caption{Electron energy spectra measured by the single channel electron spectrometer SC-ESM. Less electrons are measured with magnetic reconnection (two coils) than those shots without magnetic reconnection (one coil). No characteristic peaks seen in the OU-ESM measurements [Fig.~\ref{fig:OU_ESM}(a)] are observed in either spectra. \label{fig:SCESM}}%
\end{figure}

The electron energy spectra measured by the single channel electron spectrometer SC-ESM further supports that the direct electric field acceleration by push reconnection is likely the acceleration mechanism. In Fig.~\ref{fig:setup}(c), $E_{rec}$ in the push phase points from the front plate to the back plate. \revise{Therefore, SC-ESM measures electrons moving against the force they experience, $-eE_{rec}$. The measured electron energy spectra are shown in Fig.~\ref{fig:SCESM} for both reconnection cases with two coils and no reconnection cases with only one coil. Both cases exhibits no characteristic peaks seen in the OU-ESM measurements [Fig.~\ref{fig:OU_ESM}(a)], consistent with the expectation from the direct reconnection electric field acceleration. Furthermore, less energetic electrons are detected in the reconnection cases than the no reconnection cases, again consistent with the expectation that the directional reconnection electric field accelerates electrons away from the SC-ESM direction towards the OU-ESM direction.}

\subsection{Ion and electron acoustic waves}

The second set of experiments using micro-MRX is to study current-driven instabilities, notably ion acoustic waves (IAWs). These  can be destabilized by a large relative drift between electrons and ions at low-$\beta$ \revise{without} being subject to ion Landau damping since the ion acoustic speed, $V_S\approx \sqrt{ZT_e/M}$, is much faster than the ion thermal speed $V_i \approx\sqrt{T_i/M} $.

\begin{figure}[t]
\includegraphics[trim={0 0 8.5cm 0},clip, width=.35\textwidth]{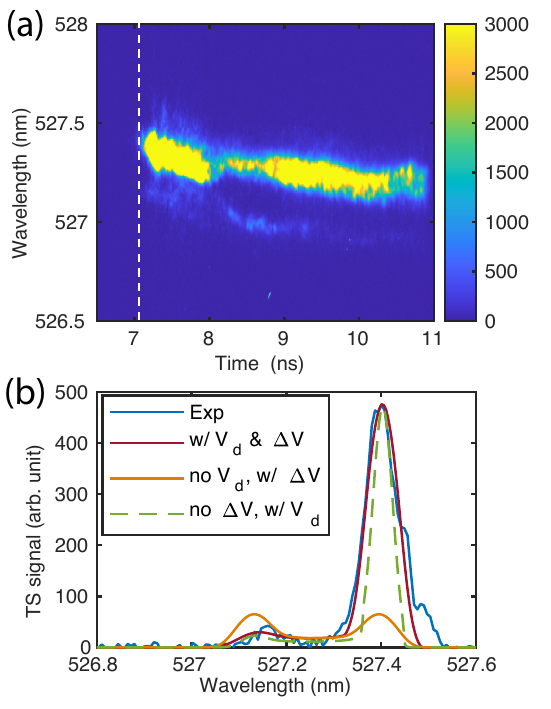}
\caption{IAW spectra (a) with linear color scale at 7--11ns from the laser irradiation. Spectrum at 7.1ns (b) along the vertical dashed line in the top panel before the IAW burst shown as blue line. Various synthetic spectra are also shown in the bottom panel. Reproduced with permission from Zhang et al., Nature Physics \textbf{19}, 909 (2023). Copyright 2023 Springer Nature.
\label{fig:IAW}}
\end{figure}

Historically, IAWs were once considered to be a promising candidate to generate anomalous resistivity~\citep{ugai77,sato79} in the reconnection diffusion region in order to realize the Petschek-like reconnection model~\citep{petschek64}. However, their importance has been dismissed due to the realization that IAWs are strongly stabilized by ion Landau damping in plasmas often found in the laboratory and in space where $T_i \simeq ZT_e$. In the micro-MRX where $T_i \ll ZT_e$, however, IAWs are not subject to ion Landau damping and can be destabilized relatively easily. Figures~\ref{fig:TStime}(a) and \ref{fig:TStime}(b) show spectra from collective Thomson scattering measured at 1--5ns from the laser irradiation and at 600 $\mu$m downstream of the central point between the top of the coils, with a propagation direction within the reconnection plane and pointing nearly (17$^\circ$ away from) downstream. The two IAW peaks in Fig. \ref{fig:TStime}(a) indicate that ion Landau damping is not large in the micro-MRX. 

Figure~\ref{fig:IAW} (a) shows IAW spectra at later times during 7--11 ns from the laser irradiation. A strong, asymmetric IAW burst, indicating large relative drift between electrons and ions, appears at $\sim$7.2 ns. The strongly asymmetric IAW spectra cannot be fitted with the theoretical spectra assuming a single Maxwellian. A spectrum is shown in Fig.~\ref{fig:IAW} (b) right before the burst at 7.1 ns along the dashed vertical line in Fig.~\ref{fig:IAW} (a). A relative drift speed between ions and electrons, $V_d$, can explain the asymmetry between two peaks (green dashed line) while ion flow gradient, $\Delta V$, can explain the broadening of each peak (orange line). Combining both $V_d=0.17 V_{th,e}$ and $\Delta V= 2\times 10^4$ m/s $\sim V_{th,i}$ can explain both features (red line). Large $V_d$ and $\Delta V$ are expected at the immediate downstream of the diffusion region where both large electric current and velocity gradients exist as part of the reconnection dynamics.

\begin{figure}[h]
\includegraphics[trim={8.5cm 0 0 0},clip, width=.35\textwidth]{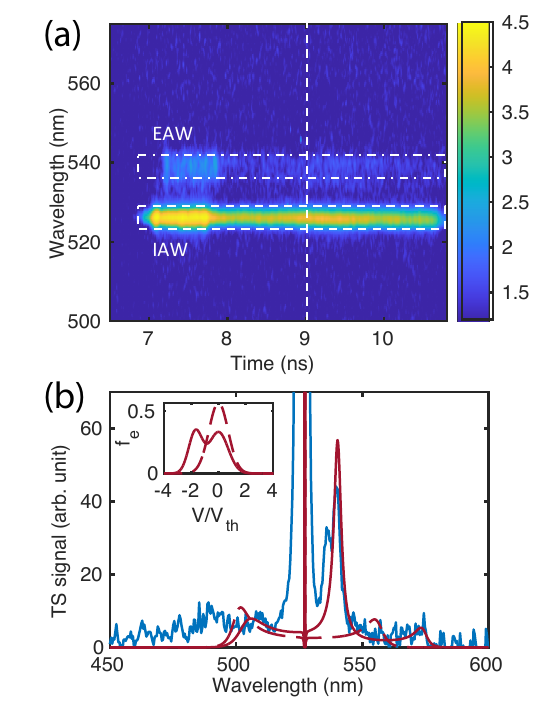}
\caption{EAW and IAW spectra (a) with logarithmic color scale at 7--11ns from the laser irradiation. Spectrum at 9ns (b) along the vertical dashed line in the top panel shown as blue line. \revise{Synthetic spectra corresponding to the Maxwellian and two-stream electron velocity distributions [inset] are shown in the dotted and solid red lines, respectively.} Reproduced with permission from Zhang et al., Nature Physics \textbf{19}, 909 (2023). Copyright 2023 Springer Nature. \label{fig:EAW}}
\end{figure}

In addition to the IAW burst, EAWs are also destabilized in micro-MRX. Figure~\ref{fig:EAW} (a) shows both IAW and EAW spectra and the latter have about 0.12 ns delay in time, implying that it is a consequence of the former. Both IAW and EAW spectra at 9 ns along the dashed vertical line are shown in Fig.~\ref{fig:EAW} (b). The strong asymmetry in the EAW cannot be explained by the synthetic spectrum (dashed red line spectrum) constructed for a single Maxwellian distribution (dashed red line in the insert) of electrons, but it can be explained by a two-stream electron distribution (solid red lines for the spectra and in the insert). Furthermore, the Doppler shift of the peak spectrum is consistent with the phase velocity of the EAWs, which is on the order of $V_{th,e}$ and matches the velocity at the valley of the distribution shown in the insert of Fig.~\ref{fig:EAW} (b).

The causality between IAW and EAW bursts are reproduced by 1D and 2D PIC simulations~\citep{zhang23}. The nonlinear evolution of IAWs in the 1D simulation shows formation of an electrostatic double layer which reflects low-energy electrons while accelerating high-energy electrons. The resultant two streams of electrons trigger EAW bursts consistent with the measurements both on the time delay between IAW and EAW bursts as well as on the EAW phase velocity. The EAW burst eventually leads to  electron heating. This scenario is also reproduced successfully in the outflow region of a 2D PIC simulation of magnetic reconnection using OSIRIS code~\citep{fonseca02} with a cold ion population in the background, albeit at a reduced mass ratio.

\section{Future prospects}\label{summary}

We have developed a unique experimental platform, the micro-MRX, to study magnetically driven reconnection at low upstream beta using high-power lasers. The platform is based on strong electric currents generated by  targets of capacitor coils, distinctly different from the reconnection experimental platform using lasers by colliding plasma plumes which are flow-driven at high upstream beta. Compared with  traditional magnetized plasma experiments studying reconnection at low upstream beta, the uniqueness of the micro-MRX platform is two-fold: (1) the \textit{ex-situ} detection capabilities of particles and photons, and (2) high-charged majority ions so that ion acoustic waves are unstable due to large electric current without ion Landau damping.

Taking advantage of this uniqueness, two initial experimental campaigns haven been successfully carried out on micro-MRX detecting  electron acceleration and excitation of ion acoustic waves by magnetic reconnection. There are ample further opportunities along each of these research lines. On the topic of electron acceleration, acceleration mechanisms other than direct acceleration by reconnection electric field can be studied also in the electron-only reconnection regime. These include Fermi acceleration by multiple plasmoids, betatron acceleration when magnetic field is compressed, and parallel electric field acceleration especially during the presence of a guide field. Each of these mechanisms can be studied with properly designed targets in the electron-only regime, even though these mechanisms were originally proposed in the electron-ion reconnection regime. For example, by having multiple coils on each side of the reconnection upstream \revise{region}, an elongated current sheet can be driven to form multiple plasmoids in favor of Fermi acceleration of electrons. This idea has been tested numerically as described below.

\begin{figure}[t]
\includegraphics[width=0.5\textwidth]{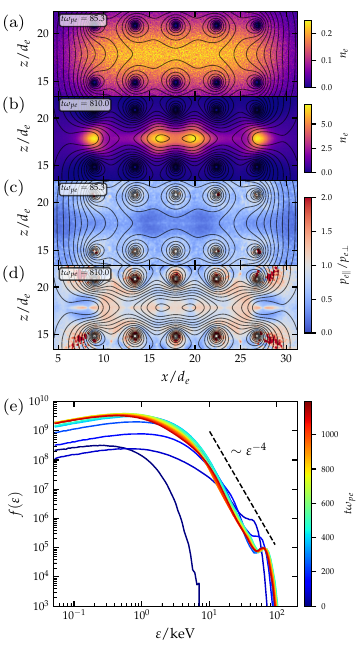}	
\caption{\label{fig:rec_hed} \ks{2D VPIC simulation of a five-coil setup. (a)--(b) Electron density at $t\omega_{pe}=85.3$ and $t\omega_{pe}=810.0$. The black lines are the contour lines of the out-of-plane component of the magnetic vector potential. (c)--(d) Electron anisotropy $p_{e\parallel}/p_{e\perp}$ at the two time frames. (e) Time evolution of the electron energy spectra.}}
\end{figure}

Figure~\ref{fig:rec_hed} shows the results from a 2D VPIC simulation of five-coil targets. The simulation has a mass ratio of 324, the same electron and ion temperature of 500 eV, ion charge state $Z=4$, and an electron density of $10^{18}$ cm$^{-3}$, resulting in an ion skin depth $d_i=0.04538$ cm and an electron skin depth $d_e=0.01$ cm. The ratio between the electron plasma frequency and the electron gyrofrequency $\omega_{pe}/\Omega_{ce}=8.43$. The simulation has a size of $L_x\times L_z=0.36$ cm $\times$ 0.36 cm =35.7$d_e\times35.7d_e$. The grid size is $564\times564$. The horizontal and vertical separations between the coils are 0.045 cm and 0.055 cm, respectively. The simulation has open boundary conditions along both directions, and the formation of the current sheet and the consequent reconnection is driven by electric current in the coils. The current increases linearly during a ramp-up time of $t_{\text{ramp}}=0.46\Omega_{ci}^{-1}=314.1\omega_{pe}^{-1}$ and then exponentially decays with a decay time of $\tau=1.97\Omega_{ci}^{-1}=1345.2\omega_{pe}^{-1}$. Initially, 1\% of all particles were loaded as background. As the simulation proceeded, five plasma injectors were included between the five pairs of coils to mimic the Gaussian laser pulses. These Gaussian injectors have a width of 0.0423 cm, and their intensity linearly increases with time until $t_{\text{ramp}}$, when the injectors are turned off. The plasma injectors result in a nearly uniform plasma density in the current sheet early in the simulation (Fig.~\ref{fig:rec_hed} (a)).

The simulation shows an elongated current sheet that thins and eventually breaks into four plasmoids (Fig.~\ref{fig:rec_hed} (a)--(b)). The simulation captures two middle islands merging, with the other two ejecting from the layer. The reconnection, occurring in an electron-only regime due to the layer's size. Early in the simulation, the acceleration was identified as  direct acceleration by the reconnection electric field~\cite{chien23}, as shown in Fig.~\ref{fig:rec_hed} (c) for an anisotropic electron distribution with $p_{e\perp}>p_{e\parallel}$, where $p_{e\perp}$ and $p_{e\parallel}$ are the perpendicular and parallel electron pressure, respectively. As the current sheet breaks into magnetic islands, the acceleration is primarily due to the Fermi mechanism~\citep{drake06,dahlin14,li17}, resulting in $p_{e\parallel}>p_{e\perp}$ at the two ends of the magnetic islands (Fig.~\ref{fig:rec_hed} (d)). During these processes, electrons are accelerated to over 100 keV, forming a significant non-thermal tail with a power-law-like distribution (Fig.~\ref{fig:rec_hed} (e)). These initial results illustrate the potential of the multi-coil targets in exploring electron acceleration mechanisms by magnetic reconnection. Further consideration and optimization will be needed in terms of target design, diagnostics, and experimental setup before implementing the ideas on specific facilities.

\begin{figure}[t]
\includegraphics[width=0.5\textwidth]{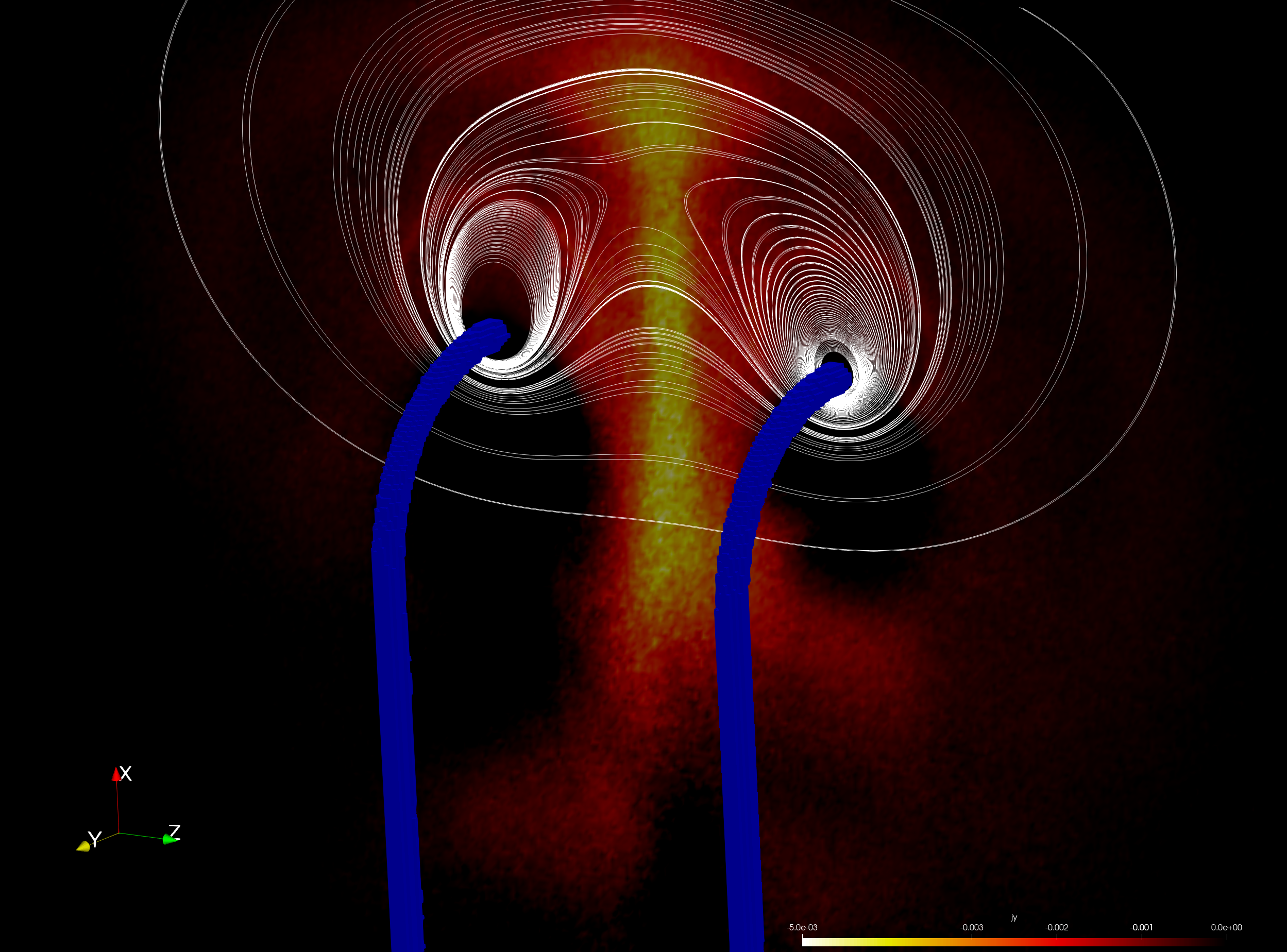}	
\caption{\label{fig:3dsim} \ks{3D PIC simulation of the micro-MRX experiment using the VPIC code. The drive coils (blue) generate magnetic field (white lines) and an electron-scale current sheet (color scale) during the push phase of reconnection at $t=0.125 \Omega_{ci}^{-1}$.}}
\end{figure}

In order to investigate the effects of a three-dimensional setup, a preliminary 3D PIC simulation of micro-MRX was carried out as shown in Fig.~\ref{fig:3dsim}. The simulation parameters are similar to those used for the multiple coil setup of Fig.~\ref{fig:rec_hed}, but using only a single pair of coils in a 3D domain. These coils (blue) contain both the straight and curved sections present in the experiment. The plasma formation via the laser interaction with the target plates is not modelled, and we instead inject plasma in a volume source between the coils during the ramp up phase, as done in Fig.~\ref{fig:rec_hed}. The grid size is $280 \times 280 \times 560$ for a simulation domain of 0.18 cm $\times$ 0.18 cm $\times$ 0.36 cm, the mass ratio is $m_i/m_e = 324$, and the separation between the coils is $0.055$ cm. Figure~\ref{fig:3dsim} shows the current density and magnetic field lines in the plane intersecting the center of the two coils at $t=0.125 \Omega_{ci}^{-1}$, early in the push phase of reconnection ($t_\textrm{ramp} = 0.46 \Omega_{ci}^{-1}$). The current sheet that forms between the two coils has a thickness comparable to the electron skin depth. Consistent with electron only reconnection~\cite{pyakurel19,stanier24lmm}, we find that the electrons are accelerated up to the electron Alfv\'en speed in bi-directional jets in the $\pm x$ outflow direction, whereas the ions remain relatively uncoupled to the magnetic field (not shown). Future studies will examine the physics of electron acceleration, as well as IAW and EAW wave generation using this 3D setup. 

On the topic of ion acoustic waves on micro-MRX, the logical next steps include quantification of the effects of waves on reconnection and detection of the waves propagating in the out-of-the-plane direction of magnetic reconnection as originally speculated to be able to facilitate the realization of Petschek reconnection~\citep{ugai77,sato79}. Other topics include extending the system size to large sizes so that ions are also coupled and ion acceleration in electron-ion reconnection can be studied. The realization of multiple X-line regimes at even larger scales and higher Lundquist numbers in the reconnection phase diagram~\citep{ji11,ji22} will allow study of the multi-scale physics of magnetic reconnection. Many major problems of magnetic reconnection research~\cite{ji20,ji23a} listed in Sec.~\ref{intro} can be studied by using the micro-MRX platform with proper targets, diagnostics and experimental setups. In this sense, comparative research with space observation by MMS~\citep{burch16}, ground-based observation by EOVSA~\citep{gary18} and STIX~\citep{krucker20}, and the upcoming FLARE experiment~\citep{ji18,ji22}, as summarized in Table~\ref{tab:table1}, will be fruitful to cover many more varieties of physics regimes and field geometry, as well as a wider parameter space.

\begin{acknowledgments}
This work was supported by the DOE Office of Science, Fusion Energy Sciences (FES) under the LaserNetUS initiative at the OMEGA Laser Facility and Jupiter Laser Facility. This work was mainly supported by the High Energy Density Laboratory Plasma Science program by the DOE Office of Science and NNSA under Grant No. DE-SC0020103. The authors express their gratitude to General Atomics, the University of Michigan, and the Laboratory for Laser Energetics for target fabrication, and to the Jupiter Laser Facility, OMEGA, and OMEGA EP crews for experimental and technical support. Work of A.M. and R.F.F was supported by the National Inertial Confinement Fusion program by the DoE NNSA at University of Rochester under contract No. DE-NA0004144.
\end{acknowledgments}

\bibliography{reconnection,1MAIN}

\end{document}